\documentclass[twocolumn]{webofc}
\usepackage[varg]{txfonts}   % Web of Conferences font
\begin{document}
\title{Atmospheric aerosol attenuation effect on FD data analysis \\ 
at the Pierre Auger Observatory}

\author{
\firstname{Laura} 
\lastname{Valore}
\inst{1}\fnsep\thanks{\email{laura.valore@na.infn.it}} %\and
        \lastname{for the Pierre Auger Collaboration}\inst{2}\fnsep\thanks{\email{auger_spokespersons@fnal.gov}}
         \fnsep\thanks{{Authors list : http://www.auger.org/archive/authors$\_$2018$\_$10.html}} %\lastname{Second author}\inst{2}\fnsep\thanks{\email{Mail address for second
            % author if necessary}} \and
        %\firstname{Third author} \lastname{Third author}\inst{3}\fnsep\thanks{\email{Mail address for last
         %  author if necessary}}
        % etc.
}

\institute{University of Naples Federico II and INFN Section of Naples, Italy
\and
       Av. San Martin Norte 304, 5613 Malarg$\rm \ddot{u}$e, Argentina    
          }

\abstract{%
  The atmospheric aerosol monitoring system of the Pierre Auger Observatory has been operating smoothly since 2004. 
  Two laser facilities (Central Laser Facility, CLF and eXtreme Laser Facility, XLF) fire sets of 50 shots four times per 
  hour during FD shifts to measure the highly variable hourly aerosol attenuation to correct the longitudinal UV light profiles 
  of the Extensive Air Showers detected by the Fluorescence Detector. Hourly aerosol attenuation loads (Vertical Aerosol Optical Depth) 
  are used to correct the measured profiles. Two techniques are used to determine the aerosol profiles, which have been proven to be 
  fully compatible. The uncertainty in the VAOD profiles measured consequently leads to an uncertainty on the energy and on the 
  estimation of the depth at the maximum development of a shower ($X_{max}$) of the event in analysis. To prove the validity of the aerosol 
  attenuation measurements used in FD event analysis, the flatness of the ratio of reconstructed SD to FD energy as a function of the 
  aerosol transmission to the depth of shower maximum has been verified.
}
\maketitle
\section{Introduction}
\label{intro}
\begin{figure}[!h]
\centering
\includegraphics[width=5cm,clip]{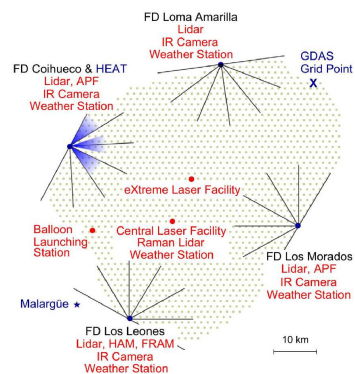}
\caption{Map of the Pierre Auger Observatory atmospheric monitoring system}
\label{fig-3}       % Give a unique label
\end{figure}

The Ultra High Energy Cosmic Rays (UHECR) primary particles entering the atmosphere produce cascades of secondary particles, 
the Extensive Air Showers (EAS). During the development of air showers, charged particles 
interact with nitrogen molecules, whose subsequent de-excitation determines the isotropic emission of fluorescence light 
(300 - 420 nm range), that provides information about the primary particle and can be detected at ground level 
with appropriate telescopes. The Pierre Auger Observatory \cite{auger} 
is the largest hybrid detector for the study of UHECR: located in Argentina, 
it has operated since 2004 with 1660 water Cherenkov detectors (Surface Detector, SD) distributed over an area of 3000 km$^2$ 
to measure the distribution of air shower particles at ground, plus 27 fluorescence telescopes (Fluorescence Detector, FD) 
overlooking the SD area, to measure the longitudinal development of air showers in the atmosphere, looking at the faint,
isotropically emitted fluorescence light they produce. While the SD has a 100$\%$ duty cycle, the FD can be operated only 
during nights with a low moon fraction and clear conditions, and this reduces its duty cycle to $14\%$.
However, the FD permits a nearly calorimetric estimate of the energy of the primary particle, while the energy estimation, when 
based on SD only measurements, must rely on hadronic interaction models. To gain all the potential of the Observatory, very 
high quality hybrid events are used to calibrate the SD energy to the FD: the importance is clear of a good measurement 
of the properties of the atmosphere, that acts as a giant calorimeter, since the flux of fluorescence photons produced is proportional 
to the energy deposited by the charged particles. The atmosphere has a double role, being responsible 
at the same time for the production and the attenuation of the fluorescence light in its travel towards the telescopes. 
It is fundamental to have a continuous monitoring of the atmospheric conditions during the data taking, with the aim of correcting 
the measurements for the real time atmospheric conditions.  The Pierre Auger Observatory carries out
an extensive program for the monitoring of the atmospheric conditions, as can be seen in figure \ref{fig-3}, where the location of
all the devices to study the atmospheric properties are shown. 

\section{The attenuation of fluorescence light}
\label{sec-2}
In the fluorescence band of nitrogen, aerosol and molecular scattering are the main mechanisms for the attenuation of light 
on its path from the point of emission to the detector. We refer to ``attenuation'' from now on, meaning the scattering of photons
out of the field of view of the detectors. 
Particular attention is given to the measurement of the aerosol attenuation profiles of UV light, the highest
variable component in space and time during the data taking. The aerosol monitoring system of the Observatory includes 
two laser facilities (the Central Laser Facility, CLF and the eXtreme Laser Facility, XLF) \cite{carlos_icrc}, two Aerosol Phase Function 
monitors (APF) \cite{apf} and the F/(Ph)otometric Robotic Aerosol Monitor (FRAM) \cite{fram_icrc}. 
Moreover, clouds are monitored with one Cloud Camera 
at each FD site \cite{cloud} and the data collected by GOES satellite \cite{goes}, lidars \cite{lidar} 
and laser facilities \cite{clfpaper} are analyzed.

The attenuation of light along a path through the atmosphere between a light source and an observer can be expressed by
a transmission coefficient $T(s)$, which gives the fraction of light not absorbed or scattered along the path $s$. 
In the UV range, the absorption of light is negligible with respect to scattering processes: the only absorber is ozone, 
which is present at higher altitudes than the atmospheric layer where air showers occur. 
\begin{figure}[!h]
\centering
\includegraphics[width=7cm,clip]{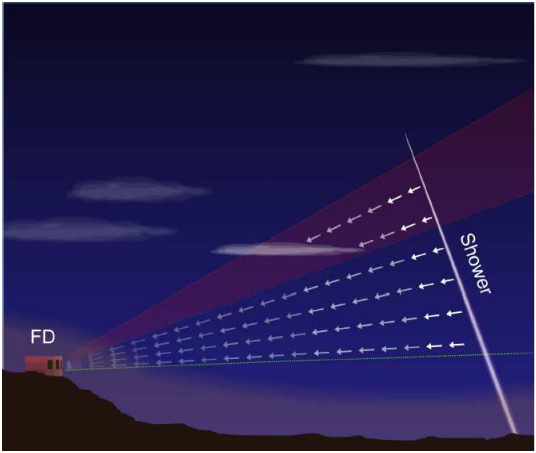}
\caption{Schematic view of the path of the light emitted in the atmosphere during the development of an air shower towards one the FD sites.}
\label{fig-10}
\end{figure}

The transmission factor $T(s)$ along the path $s$ is related to the optical depth $\tau(s)$ by the Beer-Lambert-Bouguer 
law $T(s) = \exp^{-\tau(s)}$. The aerosol and the molecular contribution can be considered separately.

The intensity of fluorescence light that reaches the FD telescopes travelling along the path $s$ from the point of emission to the 
detector depends on the transmission factors $T_{mol}$ and $T_{aer}$:
\begin{displaymath}
I(\lambda,s) = I_0(\lambda,s)\cdot T_{mol}(\lambda,s)\cdot T_{aer}(\lambda,s) \cdot (1 + H.O.) \cdot \frac{d\Omega}{4\pi} 
\end{displaymath}
where $\frac{d\Omega}{4\pi}$ is the solid angle subtended by the telescope diaphragm as seen from the light source, the molecular 
and aerosol transmission factors represents single-scattering of photons in the field of view of the telescope and the term $H.O.$ is 
a higher-order correction to the Beer-Lambert-Bouguer law that accounts for the multiple scattering of Cherenkov and fluorescence 
photons into the field of view. 

\begin{figure}[h]
\centering
\includegraphics[width=9cm,clip]{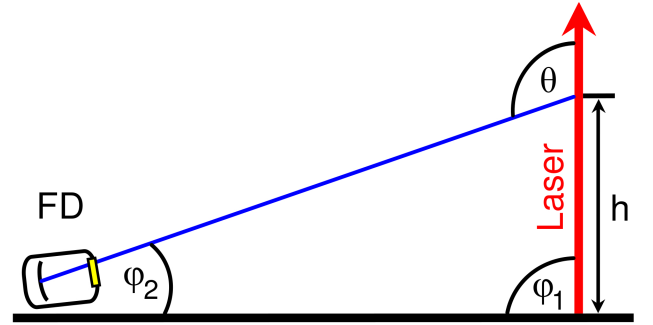}
\caption{Distant laser facility - fluorescence detector geometry.  The vertical beam at altitude h is scattered at an 
angle $\theta$ towards the FD, which corresponds to the elevation angle $\phi_2$.}
\label{fig-2}       % Give a unique label
\end{figure}

Under the assumption of a horizontally uniform atmosphere, the aerosol trasmission factor along the path $s$ can be written 
as a function of the vertical aerosol extinction coefficient $\alpha_{aer}(h)$:
\begin{displaymath}
T_{aer}(\lambda,s) =\exp \left( -\int_0^h \alpha_{aer}(z)dz/sin\phi_2 \right)
\end{displaymath}
where $\phi_2$ is the elevation angle of the light path, and since 
\begin{displaymath}
\tau_{aer}(h) = \int_0^h \alpha_{aer}(z)dz  
\end{displaymath}
the trasmission factor can be written as a function of $\tau_{aer}(h)$, also known as Vertical Aerosol Optical Depth or VAOD, as follows:
\begin{displaymath}
T_{aer}(\lambda,s) = \exp \left(-\tau_{aer}(h)/sin\phi_2\right) = \exp \left(-VAOD/sin\phi_2\right). 
\end{displaymath}
Measuring  $\tau_{aer}(h)$ at a certain altitude $h$ and knowing the elevation angle $\phi{_2}$, 
the transmission factor is calculated and can be applied as a correction to derive the intensity profile of the light at the source. 
Distant laser facilities are used to measure the aerosol optical depth at the Observatory: in figure \ref{fig-2}, the geometrical 
layout of the Pierre Auger Observatory Fluorescence Detector - vertical distant laser facility 
is schematized. 

\begin{figure}[h]
\centering
\includegraphics[width=8cm,clip]{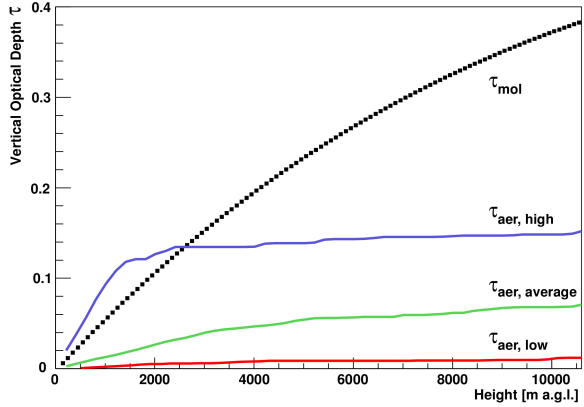}
\caption{Comparison between a typical average vertical molecular optical depth profile (black) and aerosol optical depth profiles in three different attenuation conditions (high, average and low).}
\label{fig-1}       % Give a unique label
\end{figure}

The aerosol (Mie) scattering attenuates the UV light less than molecular (Rayleigh) scattering, as displayed in figure \ref{fig-1}, where
the average molecular optical depth is compared with 3 different aerosol optical depths (high, average and low). However, it has a non negligible effect 
on the Fluorescence Detector (FD) data analysis. Of all the properties of the atmosphere, the scattering of light 
due to aerosols is the most variable phenomenon in time and space, and it influences the longitudinal development 
of the air showers: the aerosol variation with altitude, in addition to simple attenuation, modifies 
the shape of the UV light profiles, enhancing or blocking light, affecting the atmospheric depth of the maximum development of the shower or
$X_{max}$, a variable related to the nature of the primary.

Attenuation by Rayleigh scattering at the Pierre Auger Observatory is estimated using a model of the atmospheric density profile available
every three-hours via the Global Data Assimilation System, GDAS \cite{gdas}. 
Attenuation by aerosols is measured on site, and the corrections to air showers are applied using hourly aerosol 
attenuation profiles measured at each FD site \cite{clfpaper}. 
\newline

To fully determine $\tau_{aer}$, under the assumption of horizontal uniformity of the atmosphere, 
it is necessary to measure:  
\begin{itemize}
\item{the vertical height profile of the optical depth $\tau(h)$};
\item{the angular distribution of light scattered from aerosols or phase function $P(\theta)$};
\item{the wavelength dependence of the optical depth $\tau(\lambda)$}.
\end{itemize}
\subsection{Measurement of the vertical height profile of the optical depth}
\begin{figure}[h]
\centering
\includegraphics[width=8cm,clip]{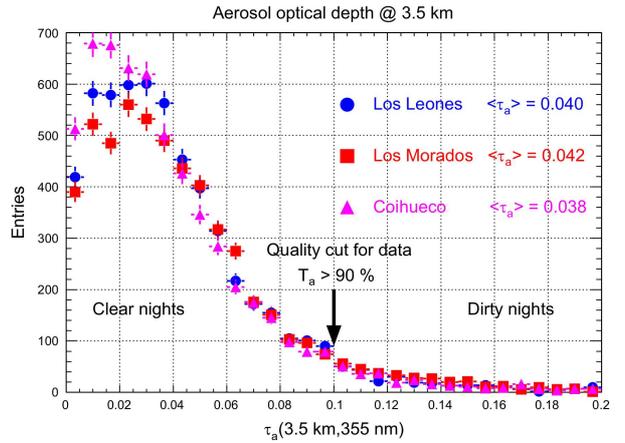}
\caption{Distribution of VAODs at 3.5 km above ground level. The average aerosol attenuations (VAODs) are highlighted in the figure.}
\label{fig-4}       % Give a unique label
\end{figure}

Two laser facilities have operated smoothly for many years: the Central Laser Facility (since 2004) and the eXtreme Laser Facility 
(since 2010) \cite{carlos_icrc}, each of them roughly equidistant from 3 out of 4 FD sites, with distances ranging from 26 to 32 km. 
Each laser facility fires sets of 50 vertical shots 4 times per hour during FD shifts.
The FD measures these UV laser tracks, and the analysis yields hourly measurements of the
aerosol attenuation loads, expressed as Vertical Aerosol Optical Depth or $\tau_{aer} (h)$
profiles. The hourly VAOD profiles are used to correct the observed longitudinal UV light
profiles of the Extensive Air Shower tracks detected by the FD. Two fully compatible analysis techniques are used to obtain the 
VAOD profiles, the Data Normalized Analysis and the Laser Simulation Analysis, both described in \cite{clfpaper}. The first one contributes
by $\sim 90\%$ to the production of the hourly aerosol attenuation profiles, the second by $\sim 10\%$ and is used as a cross check. 
Presently 11 years of hourly aerosol profiles (2004-2015) have been measured and are stored in a database accessible for air shower 
data analysis. The latest enhancements to the Data Normalized Analysis are described in \cite{max_icrc}.
In figure \ref{fig-4} the average distribution of VAODs at 3.5 km above ground level for three of the four FD sites, over the 
2004-2015 dataset, is shown. To ensure good quality events, a cut is applied to remove hours having a VAOD 
greater than 0.1, corresponding to a trasmission T > 90$\%$. The average aerosol attenuation measured at 
3.5 km above the Observatory is 0.040.
 
\subsection{Measurement of the aerosol phase function}
The angular dependence of the aerosol scattering $P(\theta)$ is described by the normalised 
differential scattering cross section in the form of the modified Henyey-Greenstein function \cite{segev}:
\begin{displaymath}
P_{aer}(\theta) = \frac{1 - g^2}{4\pi} \cdot \left( \frac{1}{(1 + g^2 - 2gcos\theta)^{3/2}} + f \frac{3cos^2\theta-1}{2(1+g^2)^{3/2}} \right)
\end{displaymath}
where the parameter $g$ measures the asymmetry of scattering, and $f$ determines the relative strength of the forward 
and backward scattering peaks. The two parameters $f$ and $g$ depend on local aerosol characteristics.
\begin{figure*}[!h]
\centering
\includegraphics[width=13cm,clip]{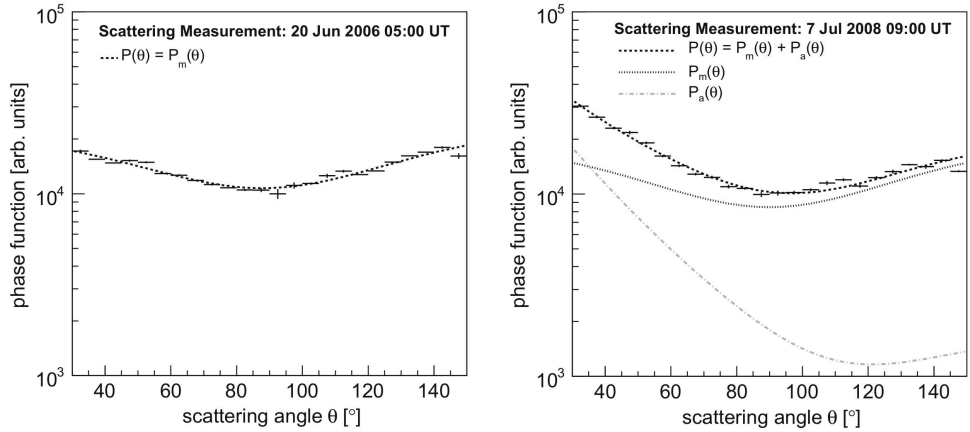}
\caption{Aerosol Phase Function (APF) light intensity vs scattering angle is shown for an aerosol-free night (left) and for a typical 
night (right).}
\label{fig-5}       % Give a unique label
\end{figure*}
Two Aerosol Phase Function (APF) monitors \cite{apf} fire collimated light pulses produced by Xenon flashers 
horizontally in the FOV of the FD sites of Los Morados and Coihueco 
and the measured data are well fitted by the parametrized phase function described above. The analysis
of the APF data lead to the values $f$ = 0.4 and $g$ = 0.6, obtained from the mean of the measured distributions over 
years 2006-2008 and confirmed by more recent data analyzed in recent years. These constant values are used in the FD data analysis. 

In figure \ref{fig-5}, examples of measurements taken with the APF monitor in Coihueco. On the left, the phase function is symmetric due to the low 
aerosol attenuation. On the right, the phase function is asymmetric, which indicates a non-negligible aerosol attenuation.

%\begin{figure}[h]
%\centering
%\includegraphics[width=8cm,clip]{fig-5.eps}
%\caption{APF.}
%\label{fig-5}       % Give a unique label
%\end{figure}

\subsection{Measurement of the wavelength dependence of the optical depth}
The wavelength dependence of the aerosol optical depth can be expressed as:
\begin{displaymath}
\tau(h,\lambda) = \tau(h,\lambda_0) \bigl( \frac{\lambda_0}{\lambda} \bigr) ^\gamma
\end{displaymath}

where $\gamma$ is the $\rm \mathring{A}$ngstrom coefficient, which was first measured using the Horizontal
Attenuation Monitor (HAM) in 2005 and confirmed by further measurements taken with the 
FRAM in the years 2006 to 2008 \cite{segev}. Both instruments are optical telescopes. The HAM is 
composed by a high intensity discharge lamp located at Coihueco plus a CCD camera positioned at Los Leones, about 45 km distant,
to observe the light across the site. The CCD camera is provided with a filter wheel to look at the source at five wavelengths 
between 350 and 550 nm. The same light source has been observed with the FRAM, an optical telescope equipped with a CCD camera 
and a photometer. 

The measured $\rm \mathring{A}$ngstrom coefficient, currently used in the air showers data analysis, is $\gamma$ = 0.7. % $\pm$ 0.5.
This value corresponds to a weak dependence of the aerosol optical depth on the wavelength, as expected for desert-like sites.

\section{How aerosols affect FD events}
The estimate of the energy deposit as a function of the slant depth from the light flux detected by the Fluorescence Detector 
is the target of the profile reconstruction, which is described in details in \cite{auger}. First of all, 
the geometry of the air shower is determined adding timing information from the surface detector to fluorescence detector data. 
Once the geometry is reconstructed, the light collected at the aperture of the FD is converted into to the energy deposit
in the atmosphere, taking into account the attenuation of the light from the point of emission to the detector and considering 
all the components that are contributing to the light profile: fluorescence light, direct Cherenkov light, aerosol (Mie) 
and molecular (Rayleigh) scattered Cherenkov light. The atmosphere is responsible for the production and
the attenuation of both the fluorescence and Cherenkov light. 

Aerosols contribute to modify the light profile as they scatter light out of the field of view during the path from the point 
of emission to the detector, resulting in an attenuation of the light at the FDs. Also, scattered Cherenkov light may enter the FOV 
due to aerosol scattering. Multiple scattering on aerosols must also be taken into account as an increase to the light profile. 
All of these contributions must be evaluated to build correctly the longitudinal energy deposit profile $dE/dX$.
Different aerosol distributions with altitude, applied as a correction to the light profile,   
lead to different modification of the shape: as an example, higher concentrations of aerosols near the ground ``push'' the $X_{max}$ 
closer to the ground. 

\begin{figure*}[!h]
\centering
\includegraphics[width=14cm,clip]{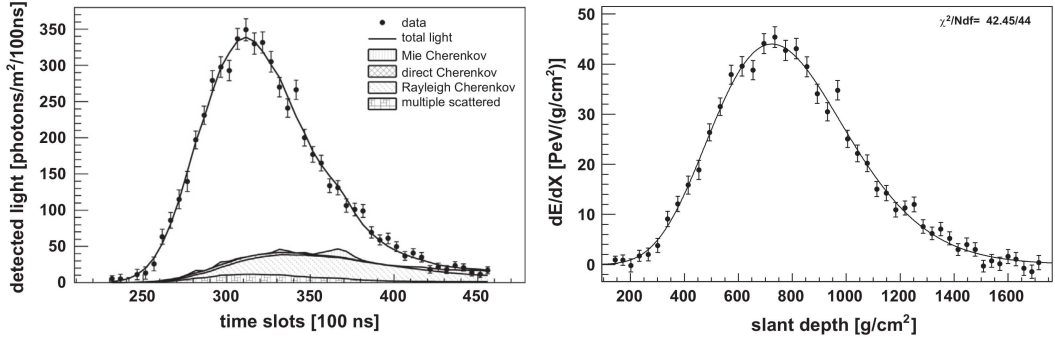}
\caption{Left: measured light profile and the separated contributions. Right: the derived energy deposit.}
\label{fig-6}       % Give a unique label
\end{figure*}

As an example of all contributions to the light profile, in figure \ref{fig-6} is shown the measured light at aperture 
of the FD together with reconstructed light contributions from each component (on the left) and the final energy 
deposit measured (on the right).

The longitudinal energy deposit profile and its maximum $\left(dE/dX \right)_{max}$ at depth $X_{max}$ are calculated 
by fitting a Gaisser-Hillas function:
\begin{displaymath}  
f_{GH}(X) = \left(dE/dX \right)_{max}\left(\frac{X - X_0}{X_{max}-X_0} \right)^{(X_{max}-X_0)/\lambda}e^{(X_{max}-X)/\lambda}
\end{displaymath}  
to the photoelectrons detected in the FDs. The energy of the air shower is calculated by integrating this equation 
and the total energy is finally obtained by correcting for the so called ``invisible energy'' carried by muons and neutrinos.
The depth of the maximum development of the profile $X_{max}$ is used in mass composition studies to determine
the nature of the primary particle. 

\subsection{Effect of aerosol uncertainties on energy and Xmax}
The laser light profiles used to measure the hourly aerosol attenuation at each FD site are affected
by uncertainties as listed in the following table. 
\newline
\newline
\begin{tabular}{lll}
\hline
 & Correlated & Uncorrelated  \\
\hline
Relative FD calibration  & 2 \% & 4 \% \\
Relative laser energy (CLF)  & 1 - 2.5 \% & 2 \% \\
Relative laser energy (XLF)  & 1 \% & 2 \% \\
Reference clean night  & 3 \% & - \\
Atmospheric variations & - & $\sim$ 3 \% \\
\hline
\label{table-1}
%\caption{Correlated and uncorrelated uncertaities to the laser light profile affecting the vertical aerosol optical depth calculation.}
\end{tabular}

The errors on the laser light propagate to the VAOD profiles and therefore to the air shower energy and $X_{max}$ 
estimation. Uncertainties are divided into correlated, meaning that they would be correlated over the 
EAS data sample, and uncorrelated from one event to the next. Both the Data Normalized and the Laser Simulation methods, 
used to estimate the VAOD profiles, are based on ratios of FD events, therefore the contribution to the uncertaities derives only from 
relative FD and laser calibrations, and not from the absolute photometric calibration. For further details, 
see \cite{laura_icrc}. 
Also, both methods are making use of a reference clean hourly profile for each year (aerosol load negligible in the hour), 
and the correlated error related to the choice of this reference profile is $3\%$. 
Finally, the uncorrelated error due to the atmospheric variations within the hour is estimated on a event-by-event basis and is $\sim 3\%$.
\begin{figure}[h!]
\centering
\includegraphics[width=8cm,clip]{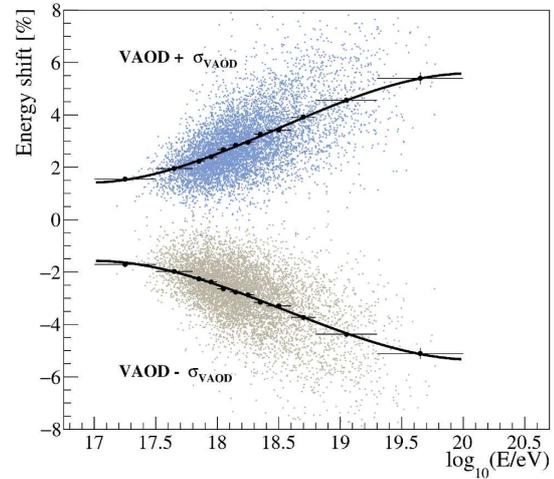}
\caption{VAOD correlated uncertainty propagated to the air shower energy estimation.}
\label{fig-7}     
\end{figure}

\begin{figure}[h!]
\centering
\includegraphics[width=8cm,clip]{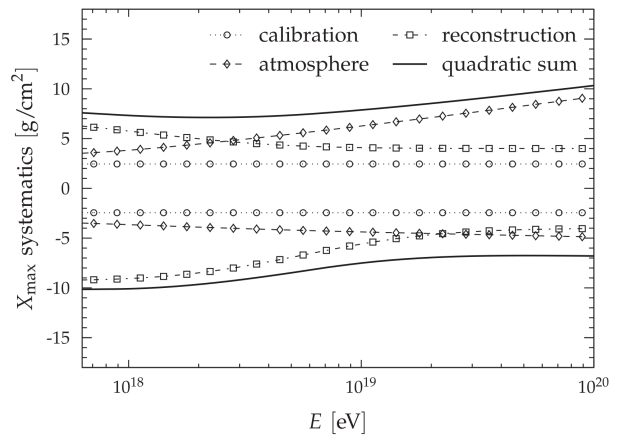}
\caption{$X_{max}$ systematic uncertainty as a function of the energy. The aerosol uncertainties are the main contribution to the ``atmosphere'' curve.}
\label{fig-8}       % Give a unique label
\end{figure}

The correlated uncertainties to the VAOD profiles lead to a systematic error to the reconstructed shower energy that is
energy dependent (as shown in figure \ref{fig-7}) and ranges from 3$\%$ to 6$\%$ from 10$^{18}$ eV to the highest energies. Also, 
the VAOD uncorrelated uncertainties lead to the same 3$\%$ - 6$\%$ to the resolution of the energy measurements.
Other contributions to the systematic error to the air showers energy are due to the systematic uncertainty of 1$\%$ 
related to the shape of the phase function and of 0.5$\%$ related to the aerosol scattering wavelength dependence \cite{valerio_icrc}. 
Concerning the effect of the aerosol uncertainties to the $X_{max}$, as shown in figure \ref{fig-8} all systematic effects due to 
the atmosphere have been combined and are energy dependent \cite{prd_xmax}. Of all the atmospheric contributions to the 
uncertainties, the aerosols are dominating, and their contribution is approximately between -4 $\rm g/cm^2$ and +8 $\rm g/cm^2$
at the highest energies. 

\subsection{Validity of the aerosol attenuation profiles applied to FD event analysis}
To confirm the validity of the aerosol attenuation profiles applied to the FD event analysis, a useful metric is the flatness of the ratio of
the reconstructed SD energy to FD energy as a function of the aerosol transmission to the shower maximum. 
On the dataset 2004-2015, with the aerosol corrections applied as described above and in \cite{max_icrc}, the slope in this ratio is 
$-0.006 \pm 0.036$, fully compatible with zero, and this is a strong indication that the aerosol attenuation profiles measured and used
in the FD data analysis of the Pierre Auger Observatory accurately describe the status of the aerosol atmosphere above the array.
\begin{figure}[h!]
\centering
\includegraphics[width=9cm,clip]{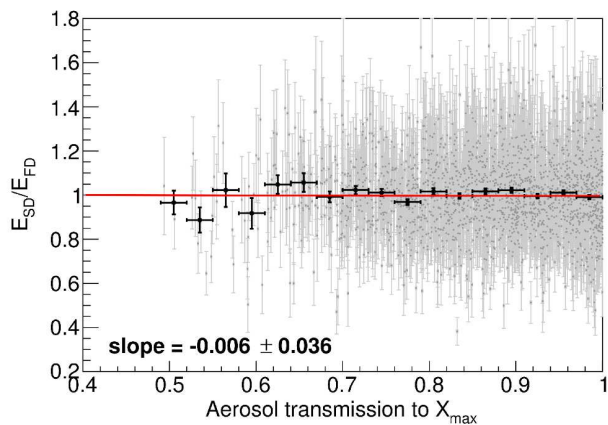}
\caption{Ratio of the SD to FD energy as a function of the aerosol transmission at $X_{max}$. The flatness proves the validity of the corrections applied.}
\label{fig-9}       % Give a unique label
\end{figure}


\begin{thebibliography}{}
%
% and use \bibitem to create references.
%
\bibitem{auger} A. Aab et al. [The Pierre Auger Collaboration], Nuclear Instruments and Methods A \textbf{798}, 172 (2015)
\bibitem{carlos_icrc} C. Medina Hernandez for the Pierre Auger Collaboration, Proceedings of the 35rd ICRC (2015), The Hague, Netherlands, arXiv:1509.03732
\bibitem{apf} S. BenZvi et al., Astroparticle Physics, Volume \textbf{28}, Issue 3, November 2007, Pages 312-320
\bibitem{fram_icrc} M. Prouza et al., Advances in Astronomy, Volume 2010, Article ID 849382
\bibitem{cloud} J. Chirinos for the Pierre Auger Collaboration, EPJ Web of Conferences \textbf{89} , 03012 (2015)
\bibitem{goes} P Abreu et al. [The Pierre Auger Collaboration], Astroparticle Physics \textbf{50-52} (2013) 92 -101
\bibitem{lidar} S. BenZvi et al., Nucl. Instrum. Methods A \textbf{574} (2007) 171 - 184
\bibitem{clfpaper} P. Abreu et al. [The Pierre Auger Collaboration], JINST \textbf{8} (2013) P04009
\bibitem{max_icrc} M. Malacari for the Pierre Auger Collaboration, Proccedings of Science, ICRC 2017, 398 (2018)
\bibitem{gdas} P. Abreu et al. [The Pierre Auger Collaboration], Astroparticle Physics \textbf{35}, 591 (2012)
\bibitem{segev} J.Abraham et al. [The Pierre Auger Collaboration], Astroparticle Physics \textbf{33}, 108 - 129 (2010)
\bibitem{laura_icrc} L. Valore for the Pierre Auger Collaboration, Proceedings of the 33rd ICRC (2013), Rio de Janeiro, Brazil p.0920 arXiv:1307.5059
\bibitem{valerio_icrc} V. Verzi for the Pierre Auger Collaboration, Proceedings of the 33rd ICRC (2013), Rio de Janeiro, Brazil p.0928 arXiv:1307.5059
\bibitem{prd_xmax} A. Aab et al. [The Pierre Auger Collaboration], Physical Review D 90, 122005 (2014)
\end{thebibliography}
\end{document}